\documentclass[aps,prb,twocolumn,groupedaddress]{revtex4-1}
\usepackage{graphicx}% Include figure files
\usepackage{epsfig}
\usepackage{dcolumn}% Align table columns on decimal point
\usepackage{bm}% Bold math
\usepackage{tabularx}% Bold math
\usepackage{hyperref}
\usepackage{xcolor}
\usepackage{amsmath}
\usepackage{float}
\usepackage{booktabs,dcolumn}
\usepackage[caption=false]{subfig} % apparently RevTeX4 is not compatible with {caption}, which is required by {subcaption}
\usepackage{textcomp}
\hypersetup{citecolor=black, filecolor= black, linkcolor= black, urlcolor= black}

\newcommand{\qvec}{$\mathbf{q}$}
\newcommand{\qe}{{\sc Quantum ESPRESSO}}
\newcommand{\qeabbrev}{{\sc QE}}

\begin{document}

% =========================
% TITLE
% =========================
\title{The pseudopotential approach within density-functional theory: the case of atomic metallic hydrogen}

% =========================
% AUTHORS & AFFILIATIONS
% =========================
\author{Jin Zhang}
%\email{zhangjin225@gmail.com}
\affiliation{Department of Physics and Astronomy, Washington State University, Pullman, WA 99164, USA}

\author{Jeffrey M. McMahon}
\email{jeffrey.mcmahon@wsu.edu}
\affiliation{Department of Physics and Astronomy, Washington State University, Pullman, WA 99164, USA}

% =========================
% ABSTRACT
% =========================
\begin{abstract}

Internal energies, enthalpies, phonon dispersion curves, and superconductivity of atomic metallic hydrogen are calculated. The (standard) use pseudopotentials in density-functional theory are compared with full (Coulomb)-potential all-electron linear muffin-tin orbital calculations. Quantitatively similar results are found as far as internal energies are concerned. Larger differences are found for phase-transition pressures; significant enough to affect the phase diagram. Electron--phonon spectral functions $\alpha^2 F(\omega)$ also show significant differences. Against expectation, the estimated superconducting critical temperature \emph{T}$_{c}$ of the first atomic metallic phase $\emph{I}4_{1}/\emph{amd}$ (Cs-IV) at $500$ GPa is actually higher.

\end{abstract}

% =========================
% MAKETITLE
% =========================
\maketitle

% =========================
% BODY OF PAPER
% =========================

% ----- ----- ----- ----- -----
\section{Introduction}
% ----- ----- ----- ----- -----

Hydrogen is the simplest and the most abundant element in the universe. Under pressure, it exhibits remarkable physics. First it solidifies and crystallizes, and then evolves through a series of high-density solid phases. In 1935, Wigner and Huntington predicted \cite{wigner1935possibility} that sufficient pressure would even dissociate hydrogen molecules, and that any Bravais lattice of such atoms would be metallic. The problem of metallic hydrogen has received considerable attention, as reviewed in Ref.\ \onlinecite{RevModPhys.84.1607}. Herein, the structures and stabilities of atomic metallic hydrogen are considered. The background of what is known (from calculations; as motivated below) and relevant to this work will be discussed in context.

Initial interest in metallic hydrogen was primarily related to astrophysical problems \cite{0034-4885-73-1-016901}. Subsequently (and more recently), there has been significant interest in it at relatively low temperatures. This can be attributed to the remarkable properties that are expected. This includes, for example, high-temperature superconductivity \cite{PhysRevLett.21.1748, PhysRevB.84.144515, PhysRevB.85.219902}. This will be considered herein. The possibility of a zero-temperature liquid ground-state has also been suggested \cite{PhysRevB.21.2641}. In this case, hydrogen may have quantum-ordered states that represent novel types of quantum fluids \cite{Babaev2004}. Applications of the (expected) remarkable physics could revolutionize several fields. Possible scientific investigations and technological uses have been speculated on in Refs.\ \onlinecite{doi:10.1080/13642819908205741, 0953-8984-29-50-504001}.

Despite experimental advances [e.g., diamond anvil cell (DAC) \cite{eremets1996high} experiments, even coupled with direct synchrotron X-ray diffraction\cite{ji2019ultrahigh}], it is still extremely difficult to measure the crystal structure of hydrogen under extreme conditions. Therefore, sophisticated calculations, often \textit{ab initio} ones based on density-functional theory (DFT) \cite{RevModPhys.87.897} have become a powerful theoretical tool to understand high-pressure hydrogen and its physical properties.

Pseudopotentials, the focus of this work, are an essential ingredient of most of these calculations. These potentials, which are smooth and nodeless, are used to replace the $1 / r$ Coulomb potential, in order to reach more rapidly convergent results. This same idea applies to the case of hydrogen, even though it only has one electron and lacks core ones.

For many properties, it is reasonable to assume that the pseudopotential should be \textit{almost} numerically identical to the Coulomb one, as long as the cutoff radius $r_{c}$ is chosen to be (extremely) small. Under high pressure, the distance between nearest-neighbor protons in atomic metallic hydrogen is approximately twofold of the Wigner--Seitz radius r$_{s}$ [$V = (4 \pi / 3) r_{s}^{3} a_{0}^{3}$, where $V$ is the volume per electron and $a_{0}$ the Bohr radius]. According to the evolution of shortest (interatomic) $\text{H}${--}$\text{H}$ distance under pressure, $r_{s}$ changes from 3.12 to 1.23 when the pressure increase from 1 atm to 500 GPa\cite{labet2012fresh}. The concern comes to be that if the pseudopotential with cutoff radius is suitable to ensure minimal core overlap.

The validity of the pseudopotential approximation in the above contexts has been discussed by McMahon and Ceperley \cite{mcmahon2011ground}. The internal energies of two structures, with Hermann--Mauguin space-group notation $I4_1 / amd$ ($c / a > 1$) (the family of structures to which this belongs will be considered further below) and $R \bar{3} m$, with different cutoff radii ($0.5$ and $0.125$ a.u.) of norm-conserving Troullier--Martins pseudopotentials \cite{PhysRevB.43.1993}, were compared. Their study indicated that this approximation has a very small impact on these calculations (subject to the above constraint). In another study \cite{geng2012high}, the energy difference between face- (\emph{Fm}$\bar{3}$\emph{m}; $fcc$) and body-centered cubic ($bcc$) phases were compared, by using a projector-augmented wave (PAW) pseudopotential \cite{blochl1994projector, PhysRevB.59.1758} and an all-electron method. This work showed that the error introduced for these calculations is insignificant. Note though that the structures considered in these studies have very high symmetry. Another important consideration is whether using a pseudopotential will influence the calculation of properties, such as the superconducting critical temperature $T_{c}$. This was made long ago by Gupta and Sinha \cite{gupta1976superconductivity}, suggesting that the estimate of $T_{c}$ may be considerably reduced by screening effects. This is based on the idea \cite{maksimov1999hydrogen} that, in the vicinity of the proton, the electron wavefunction is rigidly displaced together with the proton, and hence is not involved in the electron--phonon interaction. That is, the screening of the bare Coulomb potential should result in a decrease of coupling constant $\lambda$. This will be discussed in more detail further below.

There are still several open comments and questions concerning the use of the pseudopotential method; some specific ones are as follows: Compared with the $fcc$ and $bcc$ phases, which both belong to cubic system of crystal structures, lower-symmetry ones may be more representative and convincing. How are the internal energies of these affected? Are transition pressures (being a function of both energy and its change to first order) affected? And, is the superconductivity-physics affected?

The purpose of this work is to make a thorough analysis of the error made using pseudopotentials, using modern calculation techniques. Calculations of internal energies, (the first) phase-transition pressure, and superconducting properties of atomic metallic hydrogen under high pressures are performed. Structures that come from different crystal systems (cubic, rhombohedral, tetragonal, and orthorhombic) are considered. These quantities will be compared as calculated within the pseudopotential method to the all-electron full-potential linearised augmented-plane wave (LAPW) \cite{PhysRevB.12.3060, Koelling_1975} one.

% ----- ----- ----- ----- -----
\section{Computational Methodology}
% ----- ----- ----- ----- -----

Both the pseudopotential and all-electron calculations were performed from first principles. These were based on DFT \cite{RevModPhys.87.897}. Exchange--correlation effects were described using the generalized gradient approximation (GGA), according to the Perdew--Burke--Ernzerhof (PBE) \cite{PhysRevLett.77.3865} form. (Other) settings were chosen similarly between the two methods, for as direct comparisons as (reasonably) possible later; these are described in the following.

The pseudopotential calculations were performed using {\qe} (\qeabbrev) \cite{giannozzi2009quantum}. PAW pseudopotentials \cite{blochl1994projector, PhysRevB.59.1758} with a cutoff radius of $0.75$ a.u.\ was used to describe the region near the nucleus of hydrogen. Convergence tests (energy to within $1$ meV/proton) for this pseudopotential required $57.5$ and $345.5$ Ry for the plane-wave basis-set cutoffs (kinetic energy) for the wavefunction and charge density, respectively.

All-electron calculations were performed self-consistently using the full-potential LAPW as implemented in the Elk code \cite{elk}. A plane-wave cutoff of $\left | G + K \right |_\text{max} = 9/R_\text{min}^{MT}$ ($R_\text{min}^{MT}$ is the average of the muffin-tin radii in the unit cell) was used for the expansion of the wavefunction in the interstitial region. The muffin-tin radii for H is $0.9$ a.u.\ (comparable to that in PAW). The cutoff for charge density, which is the maximum length of $\left | G \right |$ for expanding the interstitial density was considered to $2 \left | G + K \right |_\text{max} + \varepsilon$ where $\varepsilon = 10^{{-6}}$.

It is important to briefly  recognize the difference between the all-electron (LAPW) and PAW methods. Both consider a plane-wave basis set, but augmented in the region near the nucleus to more accurately (while retaining or increasing efficiency) describe the atomic-like wavefunction. For the PAW method, however, inside the augmentation region, the (pseudo) wavefunction will be much smoother than the all-electron one. That is, the physics in this region, for this method, are similar to what happens in the pseudopotential approximation.

Convergence (to the same criterion as above) with respect to the number of $\textbf{k}$ points needed to sample (integrate over) the irreducible Brillouin zone were tested individually between {\qeabbrev} and Elk. Values obtained for the considered structures were as follows: $I4_1 / amd$ ($26^3$ both), $Cmcm$ ($26^3$ and $20^3$ for {\qeabbrev} and Elk, respectively), $I\bar{4}3d$ ($26^3$ and $28^3$), and \emph{Fm}$\bar{3}$\emph{m} ($32^3$). Smearing was used to improve convergence (of the integrations). In {\qeabbrev}, the scheme of Methfessel--Paxton \cite{PhysRevB.40.3616} was used, with a value of $0.02$ Ry; in Elk, that of Fermi--Dirac \cite{PhysRev.137.A1441}, with a suggested value \cite{elkmanual} of $0.001$ Ha.

For the phonon calculations, the GGA functional is implemented with the finite-displacement method (supercell method), but not with density-functional perturbation theory (DFPT) \cite{baroni2001phonons} in the current version of  Elk (6.3.2). To make the comparison under the same conditions, phonon dispersions were calculated using the former approach with $4{\times}4{\times}4$ supercell in both {\qeabbrev} and Elk, combined with the phonopy code\cite{togo2008first}. Note that such a grid is sufficient for a quantitative determination of the phonons in this system \cite{PhysRevB.84.144515, PhysRevB.85.219902}. For phonon dispersions, paths between high-symmetry points (covering all special points and lines necessarily and sufficiently) in the Brillouin zone were determined automatically, using the SeeK-path tool \cite{hinuma2017band}.

For the superconductivity calculations, again in order to use the GGA functional, electron--phonon coupling calculations were carried out using DFPT in {\qeabbrev}, and the supercell method in Elk. The two methods should give (numerically) the same results, as long as the sampling in reciprocal space (former method) is consistent with the supercell size (latter method); that is, the difference is one of computational efficiency \cite{RevModPhys.89.015003}. Considering this, a $4{\times}4{\times}4$ \qvec-point grid  and supercell were used for all calculations. This should be sufficient to make a quantitative \emph{comparison} between the two methods, even if only calculate approximate values of the superconducting parameters themselves \cite{PhysRevB.84.144515, PhysRevB.85.219902}.

\emph{T}$_{c}$ is estimated by numerically solving the two (complete) nonlinear Eliashberg equations. Detailed derivation of the isotropic Eliashberg gap equations have been presented by Allen and Mitrovic \cite{allen1983theory}. The following corresponding numerical method has been explained in Refs.\ \onlinecite{szczesniak2006numerical, szcze2012superconducting}. These are for the superconducting order parameter $\Delta$$_n$$\equiv$$\Delta$(\emph{i}$\omega_n$) along the imaginary frequency axis (\emph{i}=$\sqrt{-1}$), the maximum value of which corresponds to the wavefunction of the superconducting condensate, and wavefunction renormalization factor \emph{Z}$_n$$\equiv$\emph{Z}(\emph{i}$\omega_n$),
\begin{equation}\label{eq0}
{\textstyle \Delta_{n}Z_{n}=\frac{\pi}{\beta }\sum_{m=-M}^{M}\frac{\lambda (\omega_{n}-\omega_{m})-\mu ^{*}\theta (\omega_{c}-|\omega_{m}|)}{\sqrt{\omega_{m}^{2}+\Delta_{m}^{2}}}\Delta_{m}}
\end{equation}
and
\begin{equation}\label{eq1}
{\textstyle Z_{n}=1+\frac{\pi}{\beta \omega _{n}}\sum_{m=-M}^{M}\frac{\lambda (\omega _{n}-\omega _{m})}{\sqrt{\omega _{m}^{2}+\Delta _{m}^{2}}}\omega_{m}}
\end{equation}
where $\beta$=1/\emph{k}$_\text{B}$\emph{T} where \emph{k}$_\text{B}$ denotes the Boltzmann constant and $T$ the temperature, $\mu^{*}$ is the Coulomb pseudopotential, $\theta$ is the Heaviside function, $\omega_{c}$ is the phonon cut-off frequency, $\omega_{c} = 3 \omega_\text{max}$ where $\omega_\text{max}$ is the maximum phonon frequency, $\omega_{n}$=($\pi/\beta$)(2\emph{n}-1) is the $n^\text{th}$ fermion Matsubara frequency with $n = 0, {\pm}1, {\pm}2, \ldots$, the pairing kernel for electron--phonon interaction has the form $\lambda(\omega _{n} - \omega_{m}) = 2 \int_{0}^{\omega_\text{max}} d\omega ~ \frac{\alpha ^{2}F(\omega )\omega }{\omega ^{2} + (\omega _{n}-\omega _{m})^{2}}$ where $\omega$ is the phonon frequency, and $\alpha ^{2}F(\omega )$ is the Eliashberg spectral function where \emph{F}($\omega$) is the density of states of lattice vibrations (the phonon spectrum), and $\alpha^{2}$ describes the coupling of phonons to electrons on the Fermi surface. Ashcroft demonstrated \cite{PhysRevLett.78.118}, via an \textit{ab initio} calculation, that $\mu^{*} = 0.089$ in metallic hydrogen, which is similar to the (rather) standard value for a high-density system of $\mu^{*} \approx 0.1$. The former value is used herein. These two equations are solved iterative self-consistently at a certain temperature \emph{T}. \emph{T}$_{c}$ is then defined as the temperature at which the Matsubara gap $\Delta$$_n$ become zero. Herein, $2201$ Matsubara frequencies ($M = 1100$) have been used.

The most stable structures of atomic metallic hydrogen from $500$ to $3000$ GPa, as predicted by calculations, were considered. These include the $I4_1/amd$ (Cs-IV) \cite{mcmahon2011ground}, $Cmcm$ \cite{doi:10.1021/jp301596v}, and $I\bar{4}3d$ \cite{doi:10.1021/jp301596v}. Lower-symmetry, related structures, essentially the same up to a distortion(s), (such as $Fddd$ \cite{geng2012high} and $C222_1$ \cite{Geng2016} for the first two structures, respectively) were not considered; \emph{Fm}$\bar{3}$\emph{m} was also considered, for reference. The considered pressures cover the range from approximately the expected molecular-to-atomic phase transition \cite{PhysRevLett.114.105305, 2017arXiv170805217E} to just above the first predicted atomic phase transition $I4_1/amd \rightarrow Cmcm$ \cite{doi:10.1021/jp301596v}.

% ----- ----- ----- ----- -----
\section{Results and Discussion}
% ----- ----- ----- ----- -----

\begin{figure}
    \centering
    \subfloat[$c/a < 1$ ($\beta$-Sn type)]{{\includegraphics[width=3.5cm]{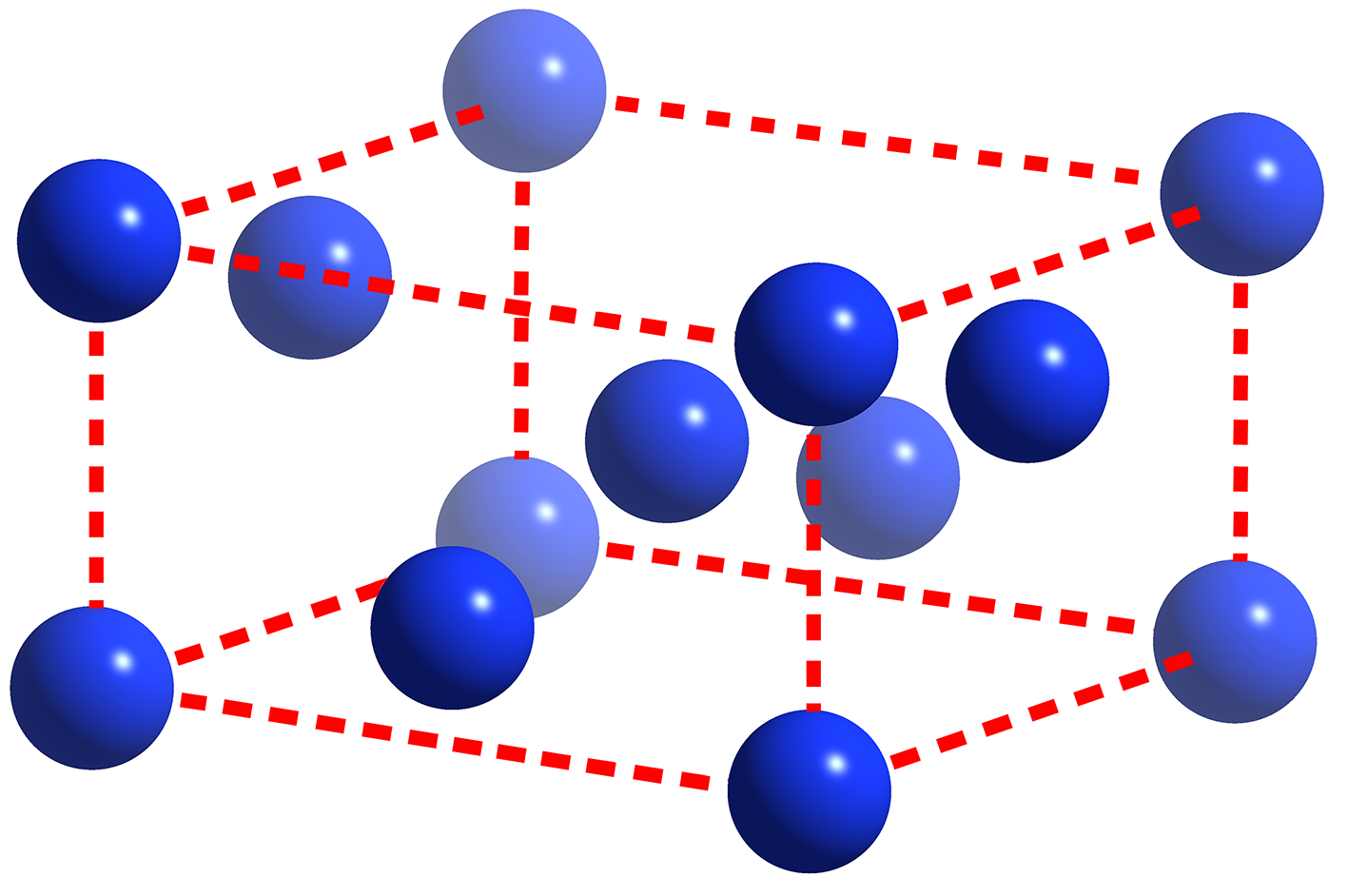}}}
    \qquad
    \subfloat[$\approx \sqrt{2}$ (diamond)]{{\includegraphics[width=5cm]{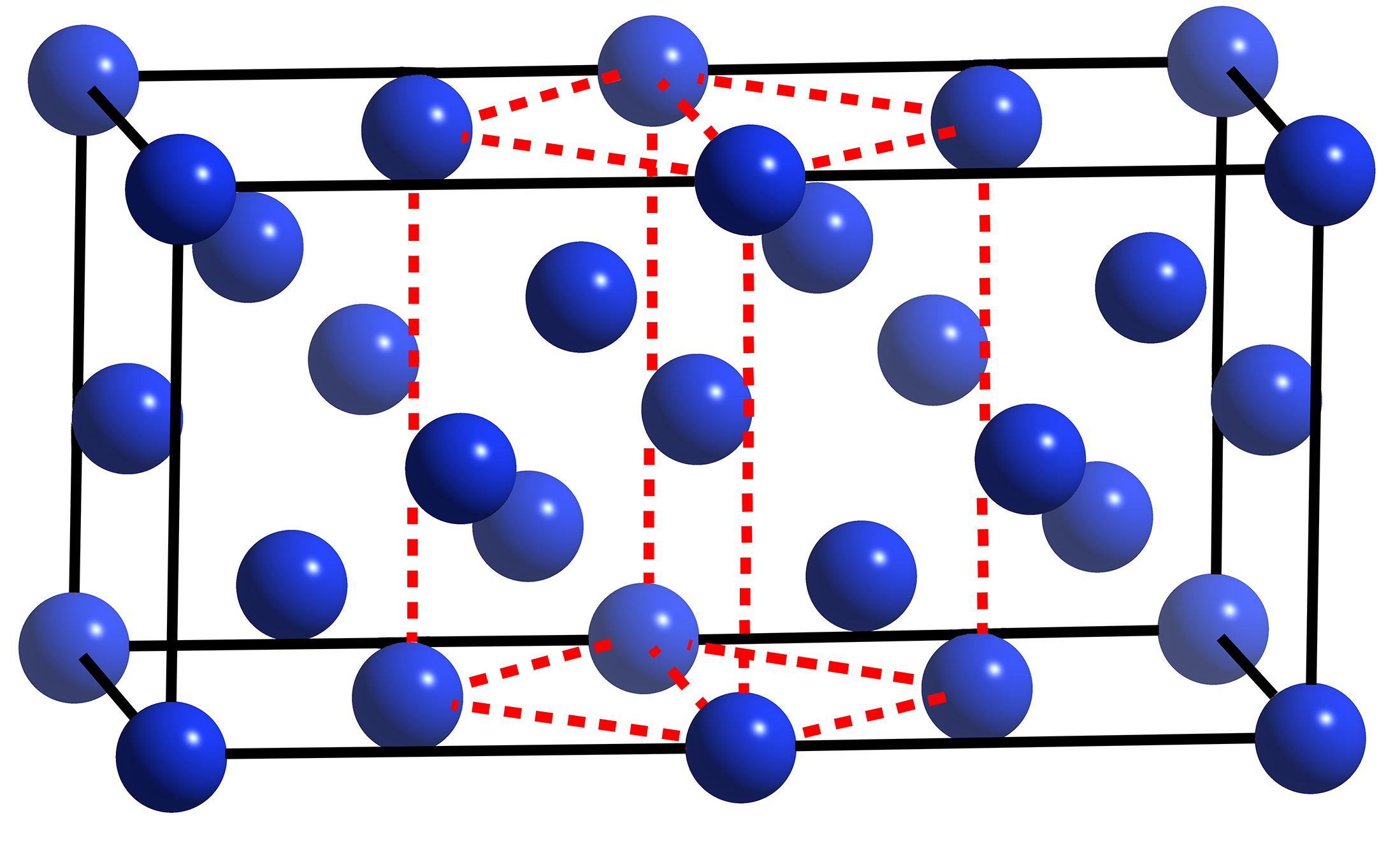}}}
    \qquad
    \subfloat[$> 1$ (Cs-IV)]{{\includegraphics[width=2.5cm]{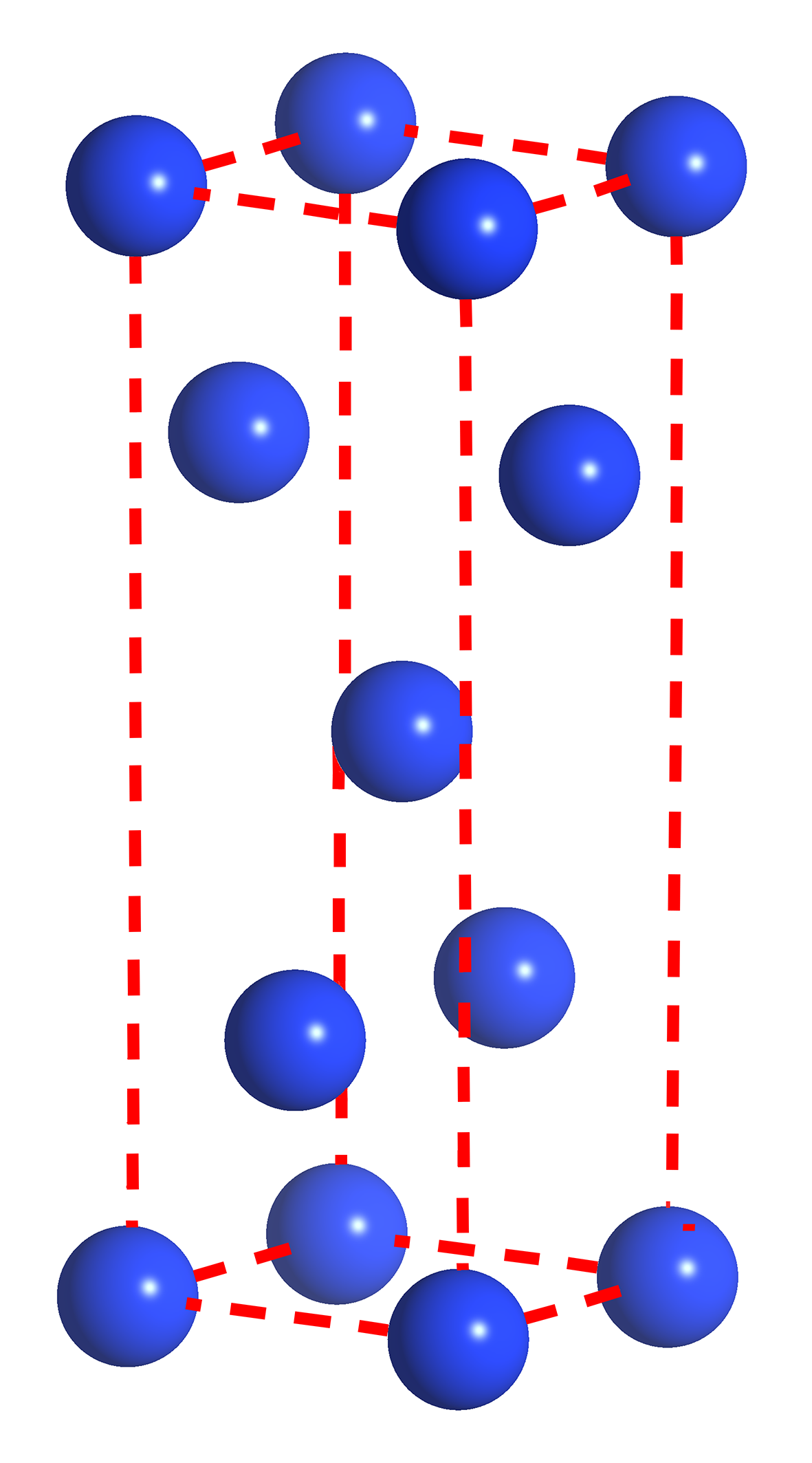}}}
    \caption{(Color online) Body-centered tetragonal (BCT) representation of (some) structures of atomic metallic hydrogen. These are characterized in terms of their $c/a$ ratio. BCT lattices are depicted in dotted red.}
    \label{fig:BCT}
\end{figure}

The structures (themselves) of high-pressure hydrogen are extremely difficult to determine by experiment. Based on first-principles calculations \cite{mcmahon2011ground}, a body-centered tetragonal (BCT) is considered to be the most promising candidate, for the first atomic phase(s). Representations of structures from this family are shown in Fig.\ \ref{fig:BCT}.
This family of structures can be characterized in terms of their $c/a$ ratio, and they are often done so using an ``elemental'' naming scheme. These are $c/a < 1$ ($\beta$-Sn type), $\approx \sqrt{2}$ (diamond), and $> 1$ (Cs-IV).

% -----
\subsection{Internal Energies}
% -----

\begin{figure}
    \centering
    \subfloat[all-electron]{{\includegraphics[width=3.75cm]{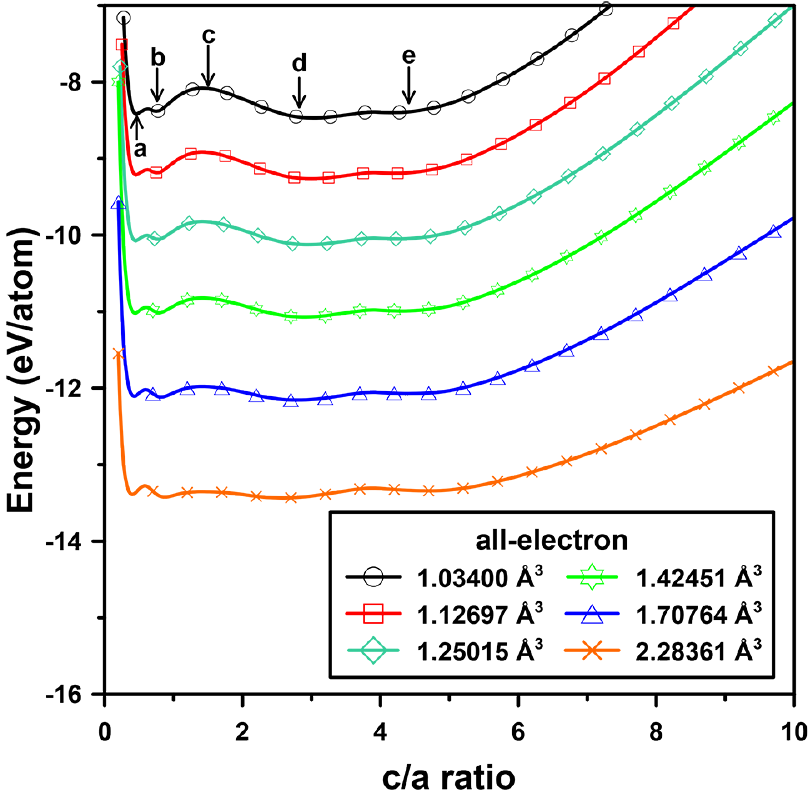}}}
    \qquad
    \subfloat[pseudopotential]{{\includegraphics[width=3.75cm]{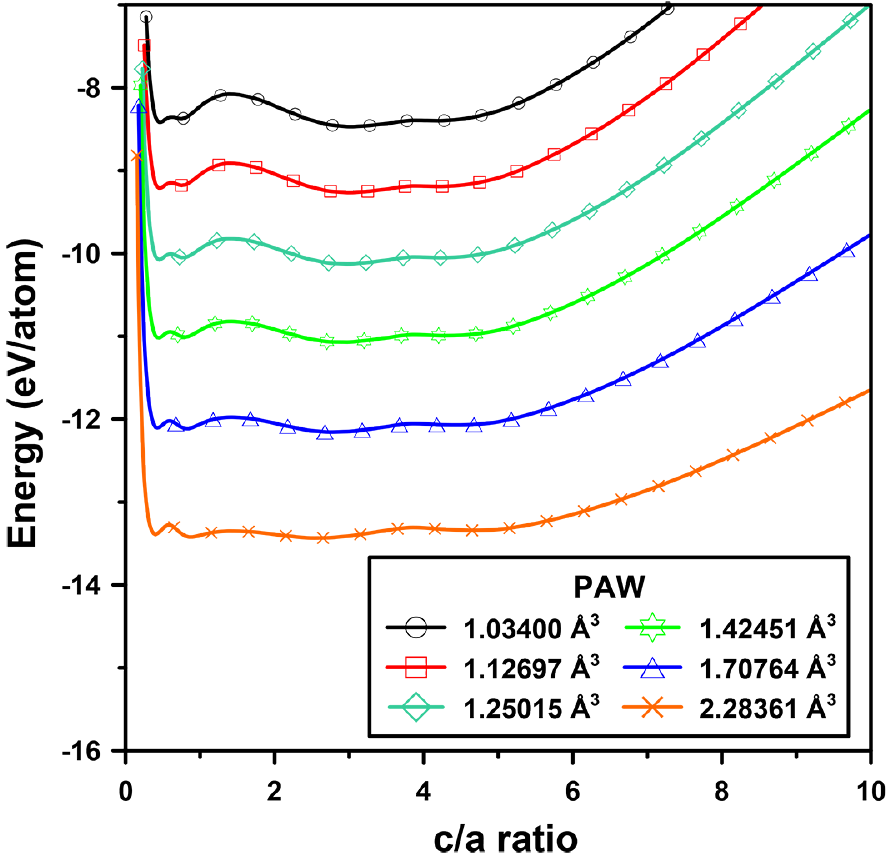}}}
    \qquad
    \subfloat[difference]{{\includegraphics[width=8cm]{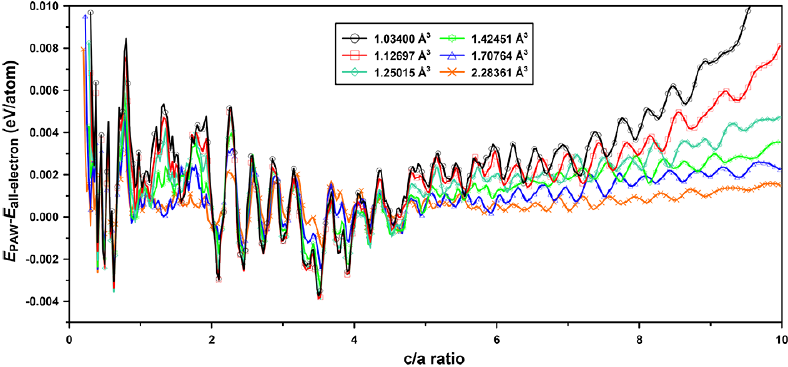}}}
    \caption{(Color online) Calculated internal energies of the BCT structures of atomic hydrogen, as a function of $c/a$ ratio at six (constant) volumes. BCT structures pointed by arrows in (a) are (from left to right): $c/a \ll 1$, $< 1$ ($\beta$-Sn), $\approx \sqrt{2}$, $> 1$ (Cs-IV), and $c/a \gg 1$, respectively. Pressures corresponding to these volumes are discussed in the text.}
    \label{fig:c_a}
\end{figure}

Internal energies as a function of $c / a$ ratio were calculated at six (constant) volumes. Note that zero-point energies were not directly included in these (or below) calculations. This ratio was varied from $0.05$ to $10$, for each volume. Volumes were determined by geometry optimizations with {\qeabbrev} over the considered pressure range (see above) in steps of $500$ GPa. These (volumes) were then fixed, and used for both {\qeabbrev} and Elk. The results are shown in Fig.\ \ref{fig:c_a}.
For both sets of calculations, there are four energy minima: a shallow one at $c/a \gg 1$, the deepest one at $c / a > 1$ (Cs-IV type), and two deep ones at $c / a < 1$ ($\beta$-Sn) and $c / a \ll 1$. Notice that $c / a \approx \sqrt{2}$ (diamond) is always unstable. From the difference plot [Fig.\ \ref{fig:c_a}(c)], a few meV/proton difference (the PAW pseudopotential energies are, in general, higher) occur on both sides of $c/a \approx 3.5$. While this difference does not change the relative stabilities of the (BCT) structures [see Fig.\ \ref{fig:c_a}(a)], it is still significant, considering the magnitude of energies.

Consider also the changes as a function of volume. The global energy minimum is always for Cs-IV. As the volume decreases, $c/a$ increases. For the pseudopotential calculations, this ranges from $2.53$ to $3.03$. For the all-electron ones, from $2.6$ to $3.05$. These ranges are in very good agreement. For both sets of calculations, the energies of $\beta$-Sn and diamond decrease with increasing volume.

The above results show that, as far as (relative) energies, structures, and both qualitative and quantitative changes with volume are concerned, the replacement of Coulomb potential by a pseudopotential appears to be reasonable. This is consistent with previous results \cite{geng2012high} focused on structures with very high symmetries. In addition (in a way) to verifying the approach, the results here extend (together, generalize) these for structures with low(er) symmetries.

% -----
\subsection{Phase Diagram}
% -----

In order to quantify the aforementioned considerations with volume, the pressure--volume ($pV$) phase diagram was constructed. This is a more sensitive measure [than internal energies (above)], as the free energy (enthalpy $H$, in this case) depends on both the energy and its first-order changes via the (hydrostatic) pressure,
\begin{equation}
\label{eq0}
    {-p} = \frac{\partial U}{\partial V}
\end{equation}
where $U$ is the internal energy. Note that pressures were calculated according to Eq.\ (\ref{eq0}); by derivatives of the equation of state (EoS) with respect to volume (instead of directly calculating the trace of external stress tensor). Specifically, once the volume dependence is known, the energy as a function of volume can be constructed, then this data is fitted with the $3^\text{rd}$-order Birch--Murnaghan EoS \cite{PhysRev.71.809}, and derivatives are calculated.

\begin{figure}
    \centering
    \subfloat[all-electron]{{\includegraphics[width=3.75cm]{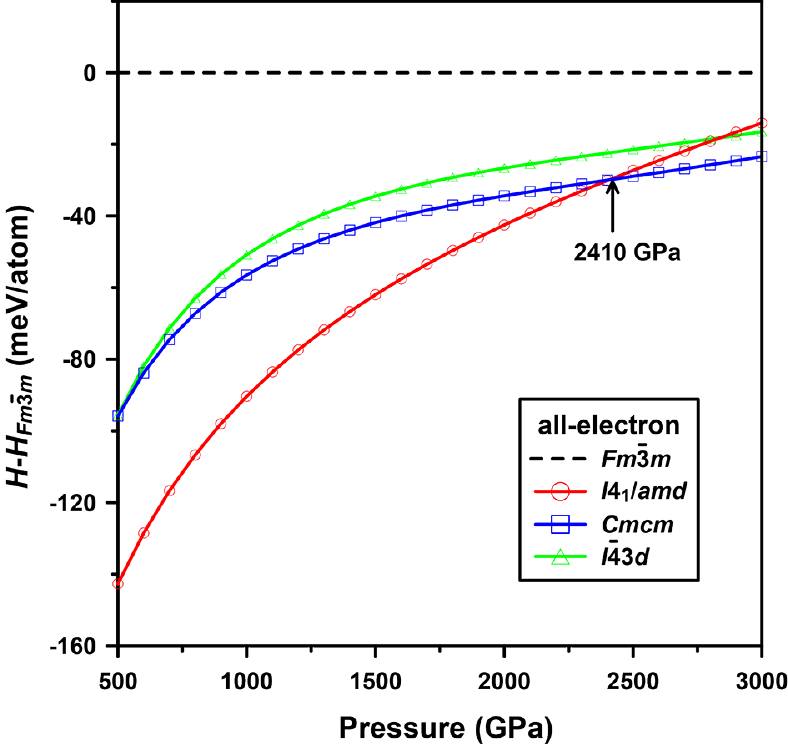}}}
    \qquad
    \subfloat[pseudopotential]{{\includegraphics[width=3.75cm]{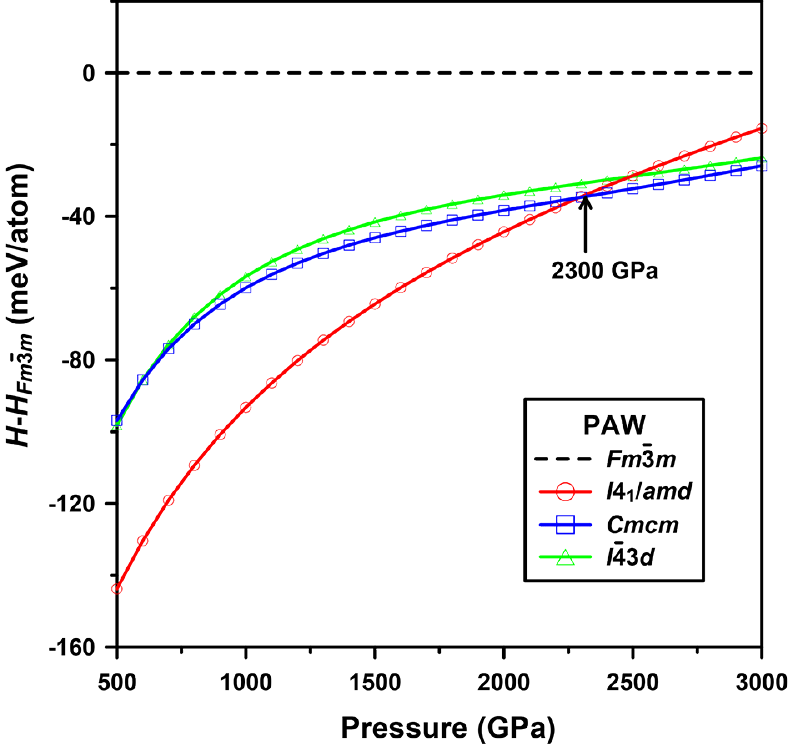}}}
    \caption{(Color online) Calculated enthalpies $H$ per atom as a function of pressure of the (predicted) most stable structures of atomic hydrogen, referenced to the \emph{Fm}$\bar{3}$\emph{m} phase.}
    \label{fig:phase_diag}
\end{figure}

Results for the first (predicted) phases of atomic hydrogen are shown in Fig.\ \ref{fig:phase_diag}.
Now using Hermann--Mauguin space-group notation (as common), these are $I4_1/amd$ (Cs-IV), $Cmcm$, and $I\bar{4}3d$. Enthalpies relative to $Fm\bar{3}m$ are shown. Note that values were calculated every $250$ GPa. The pseudopotential results are in both qualitative and quantitative agreement with earlier work \cite{doi:10.1021/jp301596v}. The all-electron ones show some important differences, however.

Consider first the trends in relative enthalpy differences. These are consistent with earlier work. In particular, $I4_1/amd$ becomes very unstable with increasing pressure, relative to a set of structures with much flatter enthalpy changes.

Consider now the phase transition pressures. That of the (first) $I4_1/amd \rightarrow Cmcm$ transition is $2300$ GPa, which is in agreement with the approximate value of $> 2100$ GPa calculated in Ref.\ \onlinecite{doi:10.1021/jp301596v} (the latter based on a less-dense pressure grid). For the all-electron calculations, the transition occurs at  $2410$ GPa. Compared to the above results (for the two pseudopotential calculations --- herein and in Ref.\ \onlinecite{doi:10.1021/jp301596v}), this difference (increase) is relatively small. But this trend appears consistent with the \textit{next} (potential) phase transition, discussed below.

Consider now the latter structures. It appears that a phase transition $Cmcm \rightarrow I\bar{4}3d$ will occur. (Indeed, but with consideration of zero-point energy. This is predicted \cite{doi:10.1021/jp301596v} above $3.5$ TPa.) Considering this next phase transition, a significant difference \textit{can} be seen. Consistent with the first transition, it appears that this one will also be pushed to even higher pressures. In this case, however, it is enough such that this transition may not occur. Consider the difference in enthalpy between these two structures, $\Delta H = H_{I\bar{4}3d} - H_{Cmcm}$. The (maximum) value with the pseudopotential approximation is $4.5$ meV/proton at $1700$ GPa; and this \emph{decreases} to $2.2$ meV/proton by $3000$ GPa. This is (even) qualitatively much different in the all-electron calculations, where $\Delta H$ \emph{increases} from $7.6$ to $7.8$ meV/proton at these pressures. That is, a phase transition, in this case, seems unlikely.

Considering the results together, all-electron calculations seem to (at least, in this region of the phase diagram considered) push transition pressures higher. Relative stabilities may also change. These results may be significant enough to change the phase diagram.

% -----
\subsection{Phonon Dispersion}
% -----

An important consideration for phase stabilities (by zero-point energy), properties (e.g., superconductivity), etc.\ is lattice vibrations. Throughout reciprocal space, these are illustrated most clearly by phonon dispersions.

\begin{figure}
\includegraphics[angle=0,width=0.9\linewidth]{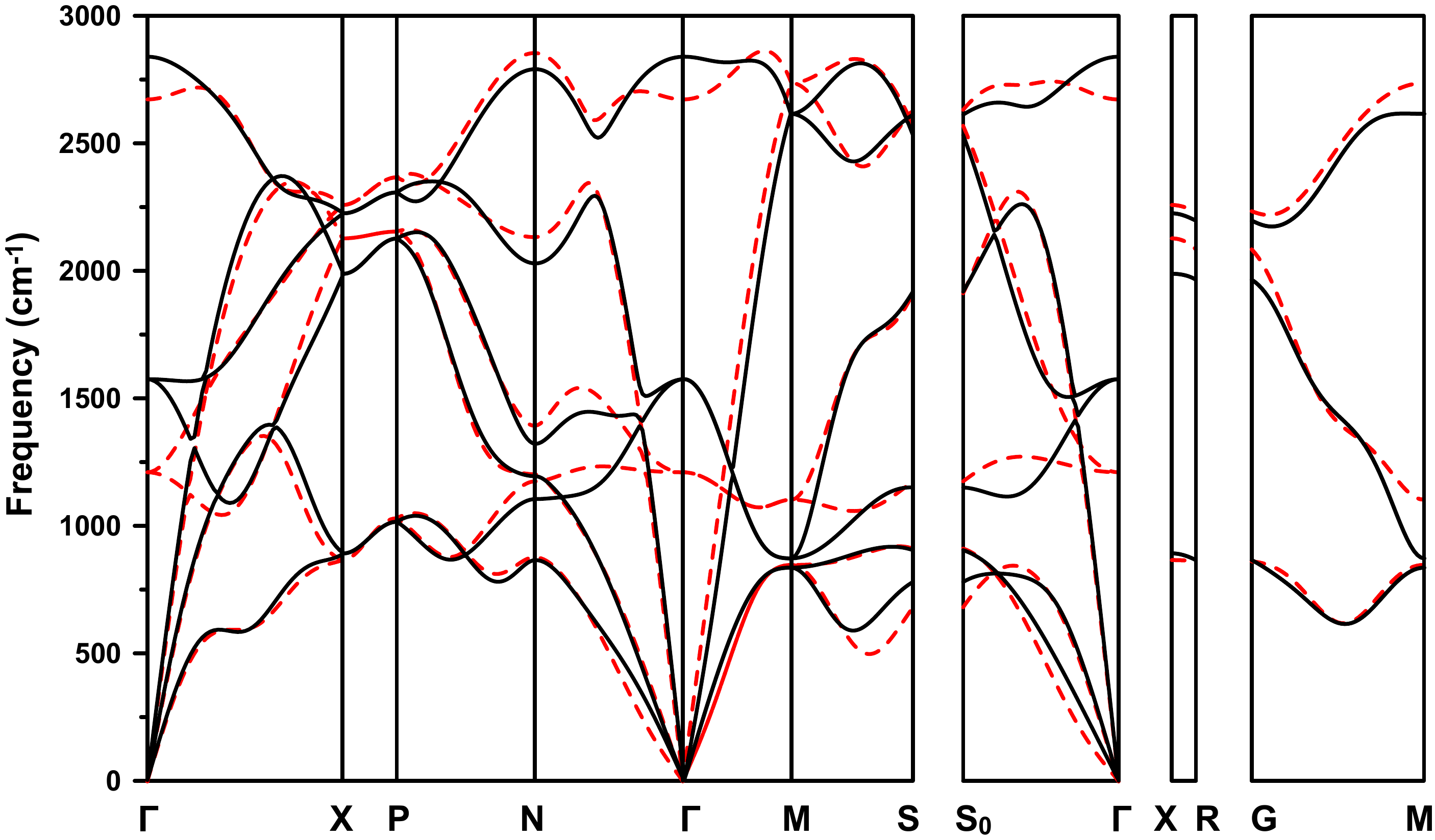}
\caption{\label{fig:phonon_dispersion} (Color online) Phonon dispersion curves for $I4_1/amd$ (Cs-IV) at $500$ GPa. Solid black curves are from the all-electron calculation, and dashed red ones from the PAW pseudopotential one.}
\end{figure}

The case of $I4_1/amd$ (Cs-IV) at $500$ GPa is considered, as an example. These results are shown in Fig.\ \ref{fig:phonon_dispersion}.
Phonon dispersion curves along various symmetry directions were calculated. Comparison shows that the dispersion relations calculated by the methods are similar. The most significant difference is near the $\Gamma$ point, where the frequencies of the optical modes as calculated with the pseudopotential approximation are much flatter. A possible explanation for this is that with the pseudopotential approximation, the electrons near the proton are not (as) bound with its motion (unlike the all-electron method --- see below); this means that the change in the electronic charge density become noticeable, and hence its phonon density of state is large and phonon dispersion flat.

% -----
\subsection{Superconductivity}
% -----

Superconductivity of atomic metallic hydrogen is considered, in this section.

\begin{figure}
\includegraphics[angle=0,width=0.6\linewidth]{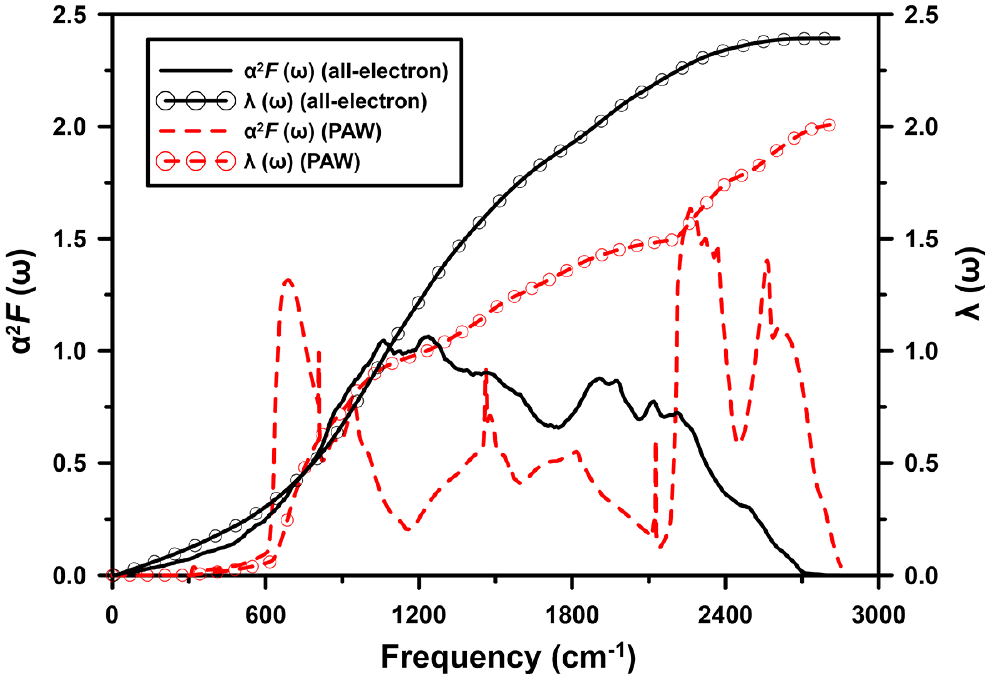}
\caption{\label{fig:Eliashberg} (Color online) Eliashberg spectral function $\alpha^{2}\emph{F}(\omega)$ and the electron--phonon coupling parameter $\lambda$ for $I4_1/amd$ at $500$ GPa.}
\end{figure}

That of $I4_1/amd$ at $500$ GPa is again used, as an example. Figure \ref{fig:Eliashberg} shows a detailed comparison of the electron--phonon spectral function $\alpha^{2}\emph{F}(\omega)$ and coupling parameter $\lambda$.
There are significant differences in the quantities as calculated by the two methods, both qualitatively and quantitatively. The pseudopotential calculations display significant (and ``peaked'') electron--phonon interaction at both (relatively) low and high frequencies, but much less at intermediate ones. This can be compared to the broad spectral function, centered at intermediate frequencies, as calculated by the all-electron method.

Only at high frequencies is this result is consistent with the work of Gupta and Sinha \cite{gupta1976superconductivity}. This interaction is mainly due to that near the proton (in metallic hydrogen). (Consider the change in the bare Coulomb interaction with $r$; this scales as $1/r^2$, and hence is largest for small $r$.) However, the electrons in this vicinity are not at all free-electron-like; their motion is bound with that of the proton. These electrons will therefore not (significantly) participate in the electron--phonon interaction.

Unlike the earlier expectation \cite{gupta1976superconductivity} of a decrease in $\lambda$, it is actually found to increase by the all-electron calculation. This can be attributed to the increase contribution to intermediate frequencies to the $\alpha^{2}\emph{F}(\omega)$.

\begin{figure}
\includegraphics[angle=0,width=0.6\linewidth]{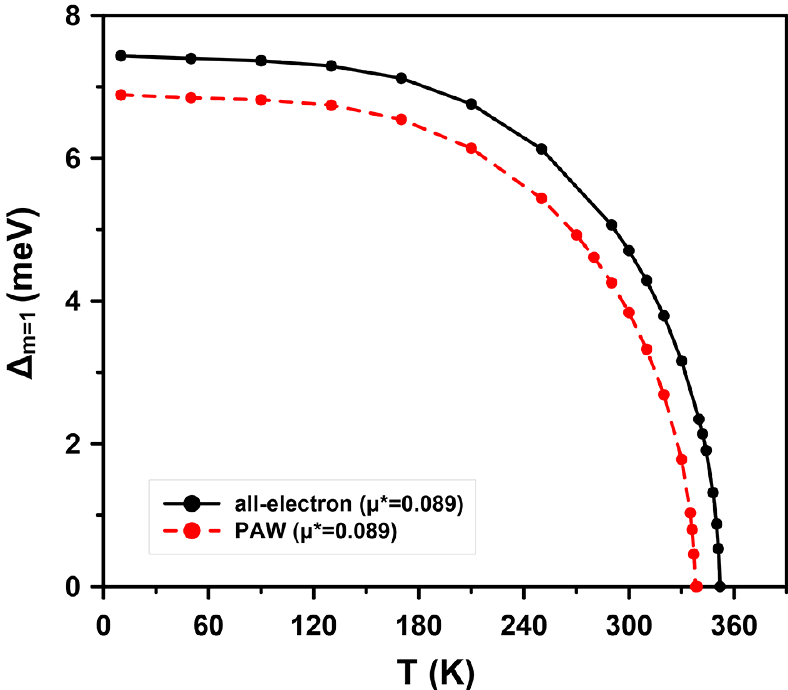}
\caption{\label{fig:Delta} (Color online) Full dependence of the maximum value of the order parameter $\Delta_{m = 1}$ on temperature for $\emph{I}4_{1}/\emph{amd}$ at $500$ GPa. Calculations are shown by both the all-electron method and with the PAW pseudopotential approximation.}
\end{figure}

The dependence of the maximum value of the order parameter $\Delta_{m = 1}$ on temperature is shown in Fig.\ \ref{fig:Delta}.
The superconducting transition temperature is defined as that at which this parameter vanishes, $\Delta_{m=1}(T_{c},\mu^{*}) = 0$. The obtained \emph{T}$_{c}$ is $352$ K in the all-electron calculation, compared to $339$ K with the pseudopotential approximation.

% ----- ----- ----- ----- -----
\section{Conclusions}
% ----- ----- ----- ----- -----

In conclusion, the reliability of (the standard use of) pseudopotentials to simulate atomic metallic hydrogen was studied. This was done for calculations of internal energy, enthalpy, phonon dispersion spectrum and superconductivity, by comparing pseudopotential to all-electron calculations. In the case of calculating internal energy, as has been considered to some extent, the accuracy that can be obtained by PAW pseudopotentials is sufficient. Differences occur for enthalpy and phonon dispersion relations, however. These may significant enough to affect the phase diagram, by both pushing transition pressures higher and changing relative stabilities. Significant differences also occur for the calculation of (at least, some) properties. For superconductivity, for example, the magnitude of the electron--phonon spectral function at both (relatively) low and high frequencies is considerably smaller as calculated by the all-electron method than with the pseudopotential approximation, while that at intermediate frequencies is increased. Together, these changes actually increase the value of $\lambda$, which causes the calculated superconducting critical temperature to be higher. These results are important for understanding metallic hydrogen; and will be so for future calculations of this system.

% ----- ----- ----- ----- -----
\section{acknowledgments}
% ----- ----- ----- ----- -----

J.\ M.\ M.\ acknowledges startup support from Washington State University and the Department of Physics and Astronomy thereat.

% ----- ----- ----- ----- -----
% REFERENCES


\begin{thebibliography}{42}%
\makeatletter
\providecommand \@ifxundefined [1]{%
 \@ifx{#1\undefined}
}%
\providecommand \@ifnum [1]{%
 \ifnum #1\expandafter \@firstoftwo
 \else \expandafter \@secondoftwo
 \fi
}%
\providecommand \@ifx [1]{%
 \ifx #1\expandafter \@firstoftwo
 \else \expandafter \@secondoftwo
 \fi
}%
\providecommand \natexlab [1]{#1}%
\providecommand \enquote  [1]{``#1''}%
\providecommand \bibnamefont  [1]{#1}%
\providecommand \bibfnamefont [1]{#1}%
\providecommand \citenamefont [1]{#1}%
\providecommand \href@noop [0]{\@secondoftwo}%
\providecommand \href [0]{\begingroup \@sanitize@url \@href}%
\providecommand \@href[1]{\@@startlink{#1}\@@href}%
\providecommand \@@href[1]{\endgroup#1\@@endlink}%
\providecommand \@sanitize@url [0]{\catcode `\\12\catcode `\$12\catcode
  `\&12\catcode `\#12\catcode `\^12\catcode `\_12\catcode `\%12\relax}%
\providecommand \@@startlink[1]{}%
\providecommand \@@endlink[0]{}%
\providecommand \url  [0]{\begingroup\@sanitize@url \@url }%
\providecommand \@url [1]{\endgroup\@href {#1}{\urlprefix }}%
\providecommand \urlprefix  [0]{URL }%
\providecommand \Eprint [0]{\href }%
\providecommand \doibase [0]{http://dx.doi.org/}%
\providecommand \selectlanguage [0]{\@gobble}%
\providecommand \bibinfo  [0]{\@secondoftwo}%
\providecommand \bibfield  [0]{\@secondoftwo}%
\providecommand \translation [1]{[#1]}%
\providecommand \BibitemOpen [0]{}%
\providecommand \bibitemStop [0]{}%
\providecommand \bibitemNoStop [0]{.\EOS\space}%
\providecommand \EOS [0]{\spacefactor3000\relax}%
\providecommand \BibitemShut  [1]{\csname bibitem#1\endcsname}%
\let\auto@bib@innerbib\@empty
%</preamble>
\bibitem [{\citenamefont {Wigner}\ and\ \citenamefont
  {Huntington}(1935)}]{wigner1935possibility}%
  \BibitemOpen
  \bibfield  {author} {\bibinfo {author} {\bibfnamefont {E.}~\bibnamefont
  {Wigner}}\ and\ \bibinfo {author} {\bibfnamefont {H.~{\'a}.}\ \bibnamefont
  {Huntington}},\ }\href@noop {} {\bibfield  {journal} {\bibinfo  {journal} {J.
  Chem. Phys}\ }\textbf {\bibinfo {volume} {3}},\ \bibinfo {pages} {764}
  (\bibinfo {year} {1935})}\BibitemShut {NoStop}%
\bibitem [{\citenamefont {McMahon}\ \emph {et~al.}(2012)\citenamefont
  {McMahon}, \citenamefont {Morales}, \citenamefont {Pierleoni},\ and\
  \citenamefont {Ceperley}}]{RevModPhys.84.1607}%
  \BibitemOpen
  \bibfield  {author} {\bibinfo {author} {\bibfnamefont {J.~M.}\ \bibnamefont
  {McMahon}}, \bibinfo {author} {\bibfnamefont {M.~A.}\ \bibnamefont
  {Morales}}, \bibinfo {author} {\bibfnamefont {C.}~\bibnamefont {Pierleoni}},
  \ and\ \bibinfo {author} {\bibfnamefont {D.~M.}\ \bibnamefont {Ceperley}},\
  }\href@noop {} {\bibfield  {journal} {\bibinfo  {journal} {Rev. Mod. Phys.}\
  }\textbf {\bibinfo {volume} {84}},\ \bibinfo {pages} {1607} (\bibinfo {year}
  {2012})}\BibitemShut {NoStop}%
\bibitem [{\citenamefont {Baraffe}\ \emph {et~al.}(2010)\citenamefont
  {Baraffe}, \citenamefont {Chabrier},\ and\ \citenamefont
  {Barman}}]{0034-4885-73-1-016901}%
  \BibitemOpen
  \bibfield  {author} {\bibinfo {author} {\bibfnamefont {I.}~\bibnamefont
  {Baraffe}}, \bibinfo {author} {\bibfnamefont {G.}~\bibnamefont {Chabrier}}, \
  and\ \bibinfo {author} {\bibfnamefont {T.}~\bibnamefont {Barman}},\
  }\href@noop {} {\bibfield  {journal} {\bibinfo  {journal} {Rep. Prog. Phys.}\
  }\textbf {\bibinfo {volume} {73}},\ \bibinfo {pages} {016901} (\bibinfo
  {year} {2010})}\BibitemShut {NoStop}%
\bibitem [{\citenamefont {Ashcroft}(1968)}]{PhysRevLett.21.1748}%
  \BibitemOpen
  \bibfield  {author} {\bibinfo {author} {\bibfnamefont {N.~W.}\ \bibnamefont
  {Ashcroft}},\ }\href@noop {} {\bibfield  {journal} {\bibinfo  {journal}
  {Phys. Rev. Lett.}\ }\textbf {\bibinfo {volume} {21}},\ \bibinfo {pages}
  {1748} (\bibinfo {year} {1968})}\BibitemShut {NoStop}%
\bibitem [{\citenamefont {McMahon}\ and\ \citenamefont
  {Ceperley}(2011{\natexlab{a}})}]{PhysRevB.84.144515}%
  \BibitemOpen
  \bibfield  {author} {\bibinfo {author} {\bibfnamefont {J.~M.}\ \bibnamefont
  {McMahon}}\ and\ \bibinfo {author} {\bibfnamefont {D.~M.}\ \bibnamefont
  {Ceperley}},\ }\href@noop {} {\bibfield  {journal} {\bibinfo  {journal}
  {Phys. Rev. B}\ }\textbf {\bibinfo {volume} {84}},\ \bibinfo {pages} {144515}
  (\bibinfo {year} {2011}{\natexlab{a}})}\BibitemShut {NoStop}%
\bibitem [{\citenamefont {McMahon}\ and\ \citenamefont
  {Ceperley}(2012)}]{PhysRevB.85.219902}%
  \BibitemOpen
  \bibfield  {author} {\bibinfo {author} {\bibfnamefont {J.~M.}\ \bibnamefont
  {McMahon}}\ and\ \bibinfo {author} {\bibfnamefont {D.~M.}\ \bibnamefont
  {Ceperley}},\ }\href@noop {} {\bibfield  {journal} {\bibinfo  {journal}
  {Phys. Rev. B}\ }\textbf {\bibinfo {volume} {85}},\ \bibinfo {pages} {219902}
  (\bibinfo {year} {2012})}\BibitemShut {NoStop}%
\bibitem [{\citenamefont {Mon}\ \emph {et~al.}(1980)\citenamefont {Mon},
  \citenamefont {Chester},\ and\ \citenamefont {Ashcroft}}]{PhysRevB.21.2641}%
  \BibitemOpen
  \bibfield  {author} {\bibinfo {author} {\bibfnamefont {K.~K.}\ \bibnamefont
  {Mon}}, \bibinfo {author} {\bibfnamefont {G.~V.}\ \bibnamefont {Chester}}, \
  and\ \bibinfo {author} {\bibfnamefont {N.~W.}\ \bibnamefont {Ashcroft}},\
  }\href@noop {} {\bibfield  {journal} {\bibinfo  {journal} {Phys. Rev. B}\
  }\textbf {\bibinfo {volume} {21}},\ \bibinfo {pages} {2641} (\bibinfo {year}
  {1980})}\BibitemShut {NoStop}%
\bibitem [{\citenamefont {Babaev}\ \emph {et~al.}(2004)\citenamefont {Babaev},
  \citenamefont {Sudbo},\ and\ \citenamefont {Ashcroft}}]{Babaev2004}%
  \BibitemOpen
  \bibfield  {author} {\bibinfo {author} {\bibfnamefont {E.}~\bibnamefont
  {Babaev}}, \bibinfo {author} {\bibfnamefont {A.}~\bibnamefont {Sudbo}}, \
  and\ \bibinfo {author} {\bibfnamefont {N.~W.}\ \bibnamefont {Ashcroft}},\
  }\href@noop {} {\bibfield  {journal} {\bibinfo  {journal} {Nature}\ }\textbf
  {\bibinfo {volume} {431}},\ \bibinfo {pages} {666} (\bibinfo {year}
  {2004})}\BibitemShut {NoStop}%
\bibitem [{\citenamefont {Nellis}()}]{doi:10.1080/13642819908205741}%
  \BibitemOpen
  \bibfield  {author} {\bibinfo {author} {\bibfnamefont {W.~J.}\ \bibnamefont
  {Nellis}},\ }\href@noop {} {\bibfield  {journal} {\bibinfo  {journal}
  {Philos. Mag. B}\ }\textbf {\bibinfo {volume} {79}},\ \bibinfo {pages}
  {655}}\BibitemShut {NoStop}%
\bibitem [{\citenamefont {Nellis}(2017)}]{0953-8984-29-50-504001}%
  \BibitemOpen
  \bibfield  {author} {\bibinfo {author} {\bibfnamefont {W.~J.}\ \bibnamefont
  {Nellis}},\ }\href@noop {} {\bibfield  {journal} {\bibinfo  {journal} {J.
  Phys. Condens. Matter}\ }\textbf {\bibinfo {volume} {29}},\ \bibinfo {pages}
  {504001} (\bibinfo {year} {2017})}\BibitemShut {NoStop}%
\bibitem [{\citenamefont {Eremets}(1996)}]{eremets1996high}%
  \BibitemOpen
  \bibfield  {author} {\bibinfo {author} {\bibfnamefont {M.}~\bibnamefont
  {Eremets}},\ }\href@noop {} {\emph {\bibinfo {title} {High Pressure
  Experimental Methods}}},\ Oxford science publications\ (\bibinfo  {publisher}
  {Oxford University Press},\ \bibinfo {year} {1996})\BibitemShut {NoStop}%
\bibitem [{\citenamefont {Ji}\ \emph {et~al.}(2019)\citenamefont {Ji},
  \citenamefont {Li}, \citenamefont {Liu}, \citenamefont {Smith}, \citenamefont
  {Majumdar}, \citenamefont {Luo}, \citenamefont {Ahuja}, \citenamefont {Shu},
  \citenamefont {Wang}, \citenamefont {Sinogeikin} \emph
  {et~al.}}]{ji2019ultrahigh}%
  \BibitemOpen
  \bibfield  {author} {\bibinfo {author} {\bibfnamefont {C.}~\bibnamefont
  {Ji}}, \bibinfo {author} {\bibfnamefont {B.}~\bibnamefont {Li}}, \bibinfo
  {author} {\bibfnamefont {W.}~\bibnamefont {Liu}}, \bibinfo {author}
  {\bibfnamefont {J.~S.}\ \bibnamefont {Smith}}, \bibinfo {author}
  {\bibfnamefont {A.}~\bibnamefont {Majumdar}}, \bibinfo {author}
  {\bibfnamefont {W.}~\bibnamefont {Luo}}, \bibinfo {author} {\bibfnamefont
  {R.}~\bibnamefont {Ahuja}}, \bibinfo {author} {\bibfnamefont
  {J.}~\bibnamefont {Shu}}, \bibinfo {author} {\bibfnamefont {J.}~\bibnamefont
  {Wang}}, \bibinfo {author} {\bibfnamefont {S.}~\bibnamefont {Sinogeikin}},
  \emph {et~al.},\ }\href@noop {} {\bibfield  {journal} {\bibinfo  {journal}
  {Nature}\ }\textbf {\bibinfo {volume} {573}},\ \bibinfo {pages} {558}
  (\bibinfo {year} {2019})}\BibitemShut {NoStop}%
\bibitem [{\citenamefont {Jones}(2015)}]{RevModPhys.87.897}%
  \BibitemOpen
  \bibfield  {author} {\bibinfo {author} {\bibfnamefont {R.~O.}\ \bibnamefont
  {Jones}},\ }\href@noop {} {\bibfield  {journal} {\bibinfo  {journal} {Rev.
  Mod. Phys.}\ }\textbf {\bibinfo {volume} {87}},\ \bibinfo {pages} {897}
  (\bibinfo {year} {2015})}\BibitemShut {NoStop}%
\bibitem [{\citenamefont {Labet}\ \emph {et~al.}(2012)\citenamefont {Labet},
  \citenamefont {Gonzalez-Morelos}, \citenamefont {Hoffmann},\ and\
  \citenamefont {Ashcroft}}]{labet2012fresh}%
  \BibitemOpen
  \bibfield  {author} {\bibinfo {author} {\bibfnamefont {V.}~\bibnamefont
  {Labet}}, \bibinfo {author} {\bibfnamefont {P.}~\bibnamefont
  {Gonzalez-Morelos}}, \bibinfo {author} {\bibfnamefont {R.}~\bibnamefont
  {Hoffmann}}, \ and\ \bibinfo {author} {\bibfnamefont {N.}~\bibnamefont
  {Ashcroft}},\ }\href@noop {} {\bibfield  {journal} {\bibinfo  {journal} {J.
  Chem. Phys.}\ }\textbf {\bibinfo {volume} {136}},\ \bibinfo {pages} {581}
  (\bibinfo {year} {2012})}\BibitemShut {NoStop}%
\bibitem [{\citenamefont {McMahon}\ and\ \citenamefont
  {Ceperley}(2011{\natexlab{b}})}]{mcmahon2011ground}%
  \BibitemOpen
  \bibfield  {author} {\bibinfo {author} {\bibfnamefont {J.~M.}\ \bibnamefont
  {McMahon}}\ and\ \bibinfo {author} {\bibfnamefont {D.~M.}\ \bibnamefont
  {Ceperley}},\ }\href@noop {} {\bibfield  {journal} {\bibinfo  {journal}
  {Phys. Rev. Lett.}\ }\textbf {\bibinfo {volume} {106}},\ \bibinfo {pages}
  {165302} (\bibinfo {year} {2011}{\natexlab{b}})}\BibitemShut {NoStop}%
\bibitem [{\citenamefont {Troullier}\ and\ \citenamefont
  {Martins}(1991)}]{PhysRevB.43.1993}%
  \BibitemOpen
  \bibfield  {author} {\bibinfo {author} {\bibfnamefont {N.}~\bibnamefont
  {Troullier}}\ and\ \bibinfo {author} {\bibfnamefont {J.~L.}\ \bibnamefont
  {Martins}},\ }\href@noop {} {\bibfield  {journal} {\bibinfo  {journal} {Phys.
  Rev. B}\ }\textbf {\bibinfo {volume} {43}},\ \bibinfo {pages} {1993}
  (\bibinfo {year} {1991})}\BibitemShut {NoStop}%
\bibitem [{\citenamefont {Geng}\ \emph {et~al.}(2012)\citenamefont {Geng},
  \citenamefont {Song}, \citenamefont {Li},\ and\ \citenamefont
  {Wu}}]{geng2012high}%
  \BibitemOpen
  \bibfield  {author} {\bibinfo {author} {\bibfnamefont {H.~Y.}\ \bibnamefont
  {Geng}}, \bibinfo {author} {\bibfnamefont {H.~X.}\ \bibnamefont {Song}},
  \bibinfo {author} {\bibfnamefont {J.}~\bibnamefont {Li}}, \ and\ \bibinfo
  {author} {\bibfnamefont {Q.}~\bibnamefont {Wu}},\ }\href@noop {} {\bibfield
  {journal} {\bibinfo  {journal} {J. Appl. Phys.}\ }\textbf {\bibinfo {volume}
  {111}},\ \bibinfo {pages} {063510} (\bibinfo {year} {2012})}\BibitemShut
  {NoStop}%
\bibitem [{\citenamefont {Bl{\"o}chl}(1994)}]{blochl1994projector}%
  \BibitemOpen
  \bibfield  {author} {\bibinfo {author} {\bibfnamefont {P.~E.}\ \bibnamefont
  {Bl{\"o}chl}},\ }\href@noop {} {\bibfield  {journal} {\bibinfo  {journal}
  {Phys. Rev. B}\ }\textbf {\bibinfo {volume} {50}},\ \bibinfo {pages} {17953}
  (\bibinfo {year} {1994})}\BibitemShut {NoStop}%
\bibitem [{\citenamefont {Kresse}\ and\ \citenamefont
  {Joubert}(1999)}]{PhysRevB.59.1758}%
  \BibitemOpen
  \bibfield  {author} {\bibinfo {author} {\bibfnamefont {G.}~\bibnamefont
  {Kresse}}\ and\ \bibinfo {author} {\bibfnamefont {D.}~\bibnamefont
  {Joubert}},\ }\href@noop {} {\bibfield  {journal} {\bibinfo  {journal} {Phys.
  Rev. B}\ }\textbf {\bibinfo {volume} {59}},\ \bibinfo {pages} {1758}
  (\bibinfo {year} {1999})}\BibitemShut {NoStop}%
\bibitem [{\citenamefont {Gupta}\ and\ \citenamefont
  {Sinha}(1976)}]{gupta1976superconductivity}%
  \BibitemOpen
  \bibfield  {author} {\bibinfo {author} {\bibfnamefont {R.}~\bibnamefont
  {Gupta}}\ and\ \bibinfo {author} {\bibfnamefont {S.}~\bibnamefont {Sinha}},\
  }in\ \href@noop {} {\emph {\bibinfo {booktitle} {Superconductivity in d-and
  f-Band Metals}}}\ (\bibinfo {organization} {Springer},\ \bibinfo {year}
  {1976})\ pp.\ \bibinfo {pages} {583--592}\BibitemShut {NoStop}%
\bibitem [{\citenamefont {Maksimov}\ and\ \citenamefont
  {Shilov}(1999)}]{maksimov1999hydrogen}%
  \BibitemOpen
  \bibfield  {author} {\bibinfo {author} {\bibfnamefont {E.~G.}\ \bibnamefont
  {Maksimov}}\ and\ \bibinfo {author} {\bibfnamefont {Y.~I.}\ \bibnamefont
  {Shilov}},\ }\href@noop {} {\bibfield  {journal} {\bibinfo  {journal}
  {Phys.-Uspekhi}\ }\textbf {\bibinfo {volume} {42}},\ \bibinfo {pages} {1121}
  (\bibinfo {year} {1999})}\BibitemShut {NoStop}%
\bibitem [{\citenamefont {Andersen}(1975)}]{PhysRevB.12.3060}%
  \BibitemOpen
  \bibfield  {author} {\bibinfo {author} {\bibfnamefont {O.~K.}\ \bibnamefont
  {Andersen}},\ }\href@noop {} {\bibfield  {journal} {\bibinfo  {journal}
  {Phys. Rev. B}\ }\textbf {\bibinfo {volume} {12}},\ \bibinfo {pages} {3060}
  (\bibinfo {year} {1975})}\BibitemShut {NoStop}%
\bibitem [{\citenamefont {Koelling}\ and\ \citenamefont
  {Arbman}(1975)}]{Koelling_1975}%
  \BibitemOpen
  \bibfield  {author} {\bibinfo {author} {\bibfnamefont {D.~D.}\ \bibnamefont
  {Koelling}}\ and\ \bibinfo {author} {\bibfnamefont {G.~O.}\ \bibnamefont
  {Arbman}},\ }\href@noop {} {\bibfield  {journal} {\bibinfo  {journal} {J.
  Phys. F Met. Phys.}\ }\textbf {\bibinfo {volume} {5}},\ \bibinfo {pages}
  {2041} (\bibinfo {year} {1975})}\BibitemShut {NoStop}%
\bibitem [{\citenamefont {Perdew}\ \emph {et~al.}(1996)\citenamefont {Perdew},
  \citenamefont {Burke},\ and\ \citenamefont
  {Ernzerhof}}]{PhysRevLett.77.3865}%
  \BibitemOpen
  \bibfield  {author} {\bibinfo {author} {\bibfnamefont {J.~P.}\ \bibnamefont
  {Perdew}}, \bibinfo {author} {\bibfnamefont {K.}~\bibnamefont {Burke}}, \
  and\ \bibinfo {author} {\bibfnamefont {M.}~\bibnamefont {Ernzerhof}},\
  }\href@noop {} {\bibfield  {journal} {\bibinfo  {journal} {Phys. Rev. Lett.}\
  }\textbf {\bibinfo {volume} {77}},\ \bibinfo {pages} {3865} (\bibinfo {year}
  {1996})}\BibitemShut {NoStop}%
\bibitem [{\citenamefont {Giannozzi}\ \emph {et~al.}(2009)\citenamefont
  {Giannozzi}, \citenamefont {Baroni}, \citenamefont {Bonini}, \citenamefont
  {Calandra}, \citenamefont {Car}, \citenamefont {Cavazzoni}, \citenamefont
  {Ceresoli}, \citenamefont {Chiarotti}, \citenamefont {Cococcioni},
  \citenamefont {Dabo} \emph {et~al.}}]{giannozzi2009quantum}%
  \BibitemOpen
  \bibfield  {author} {\bibinfo {author} {\bibfnamefont {P.}~\bibnamefont
  {Giannozzi}}, \bibinfo {author} {\bibfnamefont {S.}~\bibnamefont {Baroni}},
  \bibinfo {author} {\bibfnamefont {N.}~\bibnamefont {Bonini}}, \bibinfo
  {author} {\bibfnamefont {M.}~\bibnamefont {Calandra}}, \bibinfo {author}
  {\bibfnamefont {R.}~\bibnamefont {Car}}, \bibinfo {author} {\bibfnamefont
  {C.}~\bibnamefont {Cavazzoni}}, \bibinfo {author} {\bibfnamefont
  {D.}~\bibnamefont {Ceresoli}}, \bibinfo {author} {\bibfnamefont {G.~L.}\
  \bibnamefont {Chiarotti}}, \bibinfo {author} {\bibfnamefont {M.}~\bibnamefont
  {Cococcioni}}, \bibinfo {author} {\bibfnamefont {I.}~\bibnamefont {Dabo}},
  \emph {et~al.},\ }\href@noop {} {\bibfield  {journal} {\bibinfo  {journal}
  {J. Phys. Condens. Matter}\ }\textbf {\bibinfo {volume} {21}},\ \bibinfo
  {pages} {395502} (\bibinfo {year} {2009})}\BibitemShut {NoStop}%
\bibitem [{elk({\natexlab{a}})}]{elk}%
  \BibitemOpen
  \href@noop {} {\enquote {\bibinfo {title} {The elk code is open source,
  freely available at},}\ }\bibinfo {howpublished}
  {\url{http://elk.sourceforge.net/}} \BibitemShut {NoStop}%
\bibitem [{\citenamefont {Methfessel}\ and\ \citenamefont
  {Paxton}(1989)}]{PhysRevB.40.3616}%
  \BibitemOpen
  \bibfield  {author} {\bibinfo {author} {\bibfnamefont {M.}~\bibnamefont
  {Methfessel}}\ and\ \bibinfo {author} {\bibfnamefont {A.~T.}\ \bibnamefont
  {Paxton}},\ }\href@noop {} {\bibfield  {journal} {\bibinfo  {journal} {Phys.
  Rev. B}\ }\textbf {\bibinfo {volume} {40}},\ \bibinfo {pages} {3616}
  (\bibinfo {year} {1989})}\BibitemShut {NoStop}%
\bibitem [{\citenamefont {Mermin}(1965)}]{PhysRev.137.A1441}%
  \BibitemOpen
  \bibfield  {author} {\bibinfo {author} {\bibfnamefont {N.~D.}\ \bibnamefont
  {Mermin}},\ }\href@noop {} {\bibfield  {journal} {\bibinfo  {journal} {Phys.
  Rev.}\ }\textbf {\bibinfo {volume} {137}},\ \bibinfo {pages} {A1441}
  (\bibinfo {year} {1965})}\BibitemShut {NoStop}%
\bibitem [{elk({\natexlab{b}})}]{elkmanual}%
  \BibitemOpen
  \href@noop {} {\enquote {\bibinfo {title} {The elk code manual},}\ }\bibinfo
  {howpublished} {\url{http://elk.sourceforge.net/elk.pdf}}\BibitemShut {NoStop}%
\bibitem [{\citenamefont {Baroni}\ \emph {et~al.}(2001)\citenamefont {Baroni},
  \citenamefont {De~Gironcoli}, \citenamefont {Dal~Corso},\ and\ \citenamefont
  {Giannozzi}}]{baroni2001phonons}%
  \BibitemOpen
  \bibfield  {author} {\bibinfo {author} {\bibfnamefont {S.}~\bibnamefont
  {Baroni}}, \bibinfo {author} {\bibfnamefont {S.}~\bibnamefont
  {De~Gironcoli}}, \bibinfo {author} {\bibfnamefont {A.}~\bibnamefont
  {Dal~Corso}}, \ and\ \bibinfo {author} {\bibfnamefont {P.}~\bibnamefont
  {Giannozzi}},\ }\href@noop {} {\bibfield  {journal} {\bibinfo  {journal}
  {Rev. Mod. Phys.}\ }\textbf {\bibinfo {volume} {73}},\ \bibinfo {pages} {515}
  (\bibinfo {year} {2001})}\BibitemShut {NoStop}%
\bibitem [{\citenamefont {Togo}\ \emph {et~al.}(2008)\citenamefont {Togo},
  \citenamefont {Oba},\ and\ \citenamefont {Tanaka}}]{togo2008first}%
  \BibitemOpen
  \bibfield  {author} {\bibinfo {author} {\bibfnamefont {A.}~\bibnamefont
  {Togo}}, \bibinfo {author} {\bibfnamefont {F.}~\bibnamefont {Oba}}, \ and\
  \bibinfo {author} {\bibfnamefont {I.}~\bibnamefont {Tanaka}},\ }\href@noop {}
  {\bibfield  {journal} {\bibinfo  {journal} {Phys. Rev. B}\ }\textbf {\bibinfo
  {volume} {78}},\ \bibinfo {pages} {134106} (\bibinfo {year}
  {2008})}\BibitemShut {NoStop}%
\bibitem [{\citenamefont {Hinuma}\ \emph {et~al.}(2017)\citenamefont {Hinuma},
  \citenamefont {Pizzi}, \citenamefont {Kumagai}, \citenamefont {Oba},\ and\
  \citenamefont {Tanaka}}]{hinuma2017band}%
  \BibitemOpen
  \bibfield  {author} {\bibinfo {author} {\bibfnamefont {Y.}~\bibnamefont
  {Hinuma}}, \bibinfo {author} {\bibfnamefont {G.}~\bibnamefont {Pizzi}},
  \bibinfo {author} {\bibfnamefont {Y.}~\bibnamefont {Kumagai}}, \bibinfo
  {author} {\bibfnamefont {F.}~\bibnamefont {Oba}}, \ and\ \bibinfo {author}
  {\bibfnamefont {I.}~\bibnamefont {Tanaka}},\ }\href@noop {} {\bibfield
  {journal} {\bibinfo  {journal} {Comput. Mater. Sci.}\ }\textbf {\bibinfo
  {volume} {128}},\ \bibinfo {pages} {140} (\bibinfo {year}
  {2017})}\BibitemShut {NoStop}%
\bibitem [{\citenamefont {Giustino}(2017)}]{RevModPhys.89.015003}%
  \BibitemOpen
  \bibfield  {author} {\bibinfo {author} {\bibfnamefont {F.}~\bibnamefont
  {Giustino}},\ }\href@noop {} {\bibfield  {journal} {\bibinfo  {journal} {Rev.
  Mod. Phys.}\ }\textbf {\bibinfo {volume} {89}},\ \bibinfo {pages} {015003}
  (\bibinfo {year} {2017})}\BibitemShut {NoStop}%
\bibitem [{\citenamefont {Allen}\ and\ \citenamefont
  {Mitrovi{\'c}}(1983)}]{allen1983theory}%
  \BibitemOpen
  \bibfield  {author} {\bibinfo {author} {\bibfnamefont {P.~B.}\ \bibnamefont
  {Allen}}\ and\ \bibinfo {author} {\bibfnamefont {B.}~\bibnamefont
  {Mitrovi{\'c}}},\ }in\ \href@noop {} {\emph {\bibinfo {booktitle} {Phys.
  Status Solidi}}},\ Vol.~\bibinfo {volume} {37}\ (\bibinfo  {publisher}
  {Elsevier},\ \bibinfo {year} {1983})\ pp.\ \bibinfo {pages}
  {1--92}\BibitemShut {NoStop}%
\bibitem [{\citenamefont {Szczesniak}(2006)}]{szczesniak2006numerical}%
  \BibitemOpen
  \bibfield  {author} {\bibinfo {author} {\bibfnamefont {R.}~\bibnamefont
  {Szczesniak}},\ }\href@noop {} {\bibfield  {journal} {\bibinfo  {journal}
  {Acta Phys. Pol. A}\ }\textbf {\bibinfo {volume} {109}},\ \bibinfo {pages}
  {179} (\bibinfo {year} {2006})}\BibitemShut {NoStop}%
\bibitem [{\citenamefont {Szcze}\ \emph {et~al.}(2012)\citenamefont {Szcze},
  \citenamefont {Szcze}, \citenamefont {Drzazga} \emph
  {et~al.}}]{szcze2012superconducting}%
  \BibitemOpen
  \bibfield  {author} {\bibinfo {author} {\bibfnamefont {R.}~\bibnamefont
  {Szcze}}, \bibinfo {author} {\bibfnamefont {D.}~\bibnamefont {Szcze}},
  \bibinfo {author} {\bibfnamefont {E.}~\bibnamefont {Drzazga}},  \emph
  {et~al.},\ }\href@noop {} {\bibfield  {journal} {\bibinfo  {journal} {Solid
  State Commun.}\ }\textbf {\bibinfo {volume} {152}},\ \bibinfo {pages} {2023}
  (\bibinfo {year} {2012})}\BibitemShut {NoStop}%
\bibitem [{\citenamefont {Richardson}\ and\ \citenamefont
  {Ashcroft}(1997)}]{PhysRevLett.78.118}%
  \BibitemOpen
  \bibfield  {author} {\bibinfo {author} {\bibfnamefont {C.~F.}\ \bibnamefont
  {Richardson}}\ and\ \bibinfo {author} {\bibfnamefont {N.~W.}\ \bibnamefont
  {Ashcroft}},\ }\href@noop {} {\bibfield  {journal} {\bibinfo  {journal}
  {Phys. Rev. Lett.}\ }\textbf {\bibinfo {volume} {78}},\ \bibinfo {pages}
  {118} (\bibinfo {year} {1997})}\BibitemShut {NoStop}%
\bibitem [{\citenamefont {Liu}\ \emph {et~al.}(2012)\citenamefont {Liu},
  \citenamefont {Wang},\ and\ \citenamefont {Ma}}]{doi:10.1021/jp301596v}%
  \BibitemOpen
  \bibfield  {author} {\bibinfo {author} {\bibfnamefont {H.}~\bibnamefont
  {Liu}}, \bibinfo {author} {\bibfnamefont {H.}~\bibnamefont {Wang}}, \ and\
  \bibinfo {author} {\bibfnamefont {Y.}~\bibnamefont {Ma}},\ }\href@noop {}
  {\bibfield  {journal} {\bibinfo  {journal} {J. Phys. Chem. C}\ }\textbf
  {\bibinfo {volume} {116}},\ \bibinfo {pages} {9221} (\bibinfo {year}
  {2012})}\BibitemShut {NoStop}%
\bibitem [{\citenamefont {Geng}\ and\ \citenamefont {Wu}(2016)}]{Geng2016}%
  \BibitemOpen
  \bibfield  {author} {\bibinfo {author} {\bibfnamefont {H.~Y.}\ \bibnamefont
  {Geng}}\ and\ \bibinfo {author} {\bibfnamefont {Q.}~\bibnamefont {Wu}},\
  }\href@noop {} {\bibfield  {journal} {\bibinfo  {journal} {Sci. Rep.}\
  }\textbf {\bibinfo {volume} {6}},\ \bibinfo {pages} {36745} (\bibinfo {year}
  {2016})}\BibitemShut {NoStop}%
\bibitem [{\citenamefont {McMinis}\ \emph {et~al.}(2015)\citenamefont
  {McMinis}, \citenamefont {Clay}, \citenamefont {Lee},\ and\ \citenamefont
  {Morales}}]{PhysRevLett.114.105305}%
  \BibitemOpen
  \bibfield  {author} {\bibinfo {author} {\bibfnamefont {J.}~\bibnamefont
  {McMinis}}, \bibinfo {author} {\bibfnamefont {R.~C.}\ \bibnamefont {Clay}},
  \bibinfo {author} {\bibfnamefont {D.}~\bibnamefont {Lee}}, \ and\ \bibinfo
  {author} {\bibfnamefont {M.~A.}\ \bibnamefont {Morales}},\ }\href@noop {}
  {\bibfield  {journal} {\bibinfo  {journal} {Phys. Rev. Lett.}\ }\textbf
  {\bibinfo {volume} {114}},\ \bibinfo {pages} {105305} (\bibinfo {year}
  {2015})}\BibitemShut {NoStop}%
\bibitem [{\citenamefont {{Eremets}}\ \emph {et~al.}(2017)\citenamefont
  {{Eremets}}, \citenamefont {{Drozdov}}, \citenamefont {{Kong}},\ and\
  \citenamefont {{Wang}}}]{2017arXiv170805217E}%
  \BibitemOpen
  \bibfield  {author} {\bibinfo {author} {\bibfnamefont {M.~I.}\ \bibnamefont
  {{Eremets}}}, \bibinfo {author} {\bibfnamefont {A.~P.}\ \bibnamefont
  {{Drozdov}}}, \bibinfo {author} {\bibfnamefont {P.~P.}\ \bibnamefont
  {{Kong}}}, \ and\ \bibinfo {author} {\bibfnamefont {H.}~\bibnamefont
  {{Wang}}},\ }\href@noop {} {\bibfield  {journal} {\bibinfo  {journal} {ArXiv
  e-prints}\ } (\bibinfo {year} {2017})},\ \Eprint
  {http://arxiv.org/abs/1708.05217} {arXiv:1708.05217} \BibitemShut {NoStop}%
\bibitem [{\citenamefont {Birch}(1947)}]{PhysRev.71.809}%
  \BibitemOpen
  \bibfield  {author} {\bibinfo {author} {\bibfnamefont {F.}~\bibnamefont
  {Birch}},\ }\href@noop {} {\bibfield  {journal} {\bibinfo  {journal} {Phys.
  Rev.}\ }\textbf {\bibinfo {volume} {71}},\ \bibinfo {pages} {809} (\bibinfo
  {year} {1947})}\BibitemShut {NoStop}%
\end{thebibliography}
\end{document}